\documentstyle[12pt]{article}
\def\cmm2{{\,\rm cm^{-2}}}
\def\cm2{{\,{\rm cm}^2}}
\def\cmm3{{\,{\rm cm}^{-3}}}
\def\gcmm3{{\,{\rm g\,cm^{-3}}}}

\def\fun#1#2{\lower3.6pt\vbox{\baselineskip0pt\lineskip.9pt
  \ialign{$\mathsurround=0pt#1\hfil##\hfil$\crcr#2\crcr\sim\crcr}}}

\begin{document}
\baselineskip=18pt
\pagestyle{empty}
\begin{center}
\bigskip

\rightline{CWRU-P6-97}
\rightline{CERN-TH-97/122}
\rightline{astro-ph/9706227}

\vspace{0.5in}
{\Large \bf THE END OF THE AGE PROBLEM, AND THE CASE FOR A COSMOLOGICAL CONSTANT
REVISITED}
\vspace{0.3in}

\vspace{.2in}
Lawrence M. Krauss$^{1,2}$ \\
\vspace{.2in}
{\it $^1$Theory Division, CERN, CH-1211, Geneva 23, Switzerland\\}
\vspace{0.1in}
{\it $^2$Departments of Physics and Astronomy\\
Case Western Reserve University\\
Cleveland, OH~~44106-7079}\\
\vspace{0.1in}
(submitted to {\it Science})

\end{center}

\vspace{.3in}

\begin{abstract}

The lower
limit on the age of the universe derived from globular cluster dating
techniques, which previously strongly motivated a non-zero cosmological
constant, has now been dramatically reduced, allowing consistency for a flat
matter dominated universe with a Hubble Constant, $H_0 \le 66 km
s^{-1} Mpc^{-1}$.  The case for an open universe versus a flat universe
with non-zero
cosmological constant is reanalyzed in this context, incorporating not only the
new age data, but also updates on baryon abundance constraints, and large scale
structure arguments.  For the first time, the allowed parameter space for
the density of non-relativistic matter appears larger for an open
universe than for a flat universe with cosmological constant, while a flat
universe with zero cosmological constant remains strongly disfavored. 
Several other preliminary observations suggest a non-zero cosmological
constant, but a definitive determination awaits refined measurements of $q_0$,
and small scale anisotropies of the Cosmic Microwave background.  I argue that
fundamental theoretical arguments favor a non-zero cosmological
constant over an open universe. However, if either case is confirmed,
the challenges posed for fundamental particle physics will be great.
\end{abstract}

\newpage
\pagestyle{plain}

\baselineskip=21pt

The cosmological model perhaps most strongly favored by the data
over the past few years has involved a proposal which is heretical from an
elementary particle physics perspective.  In order to reconcile a flat
universe---favored by both inflationary models and by the longstanding
flatness problem in cosmology---with the apparent age of globular clusters,
and the fact that many estimates for the clustered mass density on large
scales suggest that insufficient non-relativistic matter exists to achieve a
flat universe, the idea that the cosmological
constant is non-zero has been invoked\cite{tsk}.  Most
recently the inclusion of additional arguments associated with the baryon
density of the Universe and large scale structure have further
strengthened the case for a cosmological constant
\cite{kraussturn,oststein}.

The problem with this from a fundamental perspective is that a cosmological
constant----associated in modern parlance with a non-zero vacuum energy density
in the universe---on a scale which would be cosmologically relevant and yet 
still allowed today would take a value which is roughly 125 orders
of magnitude smaller than the naive value one might expect based on
considerations of quantum mechanics and gravity (see for example
\cite{wein}).  This apparent discrepancy would involve the most extreme
fine tuning problem known in physics, and for this reason many particle
physicists would prefer any mechanism which would drive the cosmological
constant to be exactly zero today. 

The possibility that cosmology might force physicists to have to directly
confront this longstanding issue in fundamental physics is exciting,
but at the same time its potential significance warrants a careful and
continued examination of the data which motivates the confrontation.  In
this regard, the recent reanalysis of globular cluster age estimates based in
part on new parallax measurements obtained from the Hipparcos satellite is very
significant.

Globular Cluster ages are obtained by fitting the observed
color-magnitude diagram to the predicted distribution
for a system of stars of different masses which all form at the same time
and which are then evolved on a computer to a certain age.  A number of
different fitting techniques are employed, with different uncertainties.
Because theoretical models predict the instrinsic stellar luminosity, while
measurements yield an apparent magnitude, in order to compare theory and
prediction, the distance to the globular cluster must be known.  Using
an extensive Monte Carlo analysis approach begun several years
ago\cite{chabetal} it was demonstrated that it is precisely this distance
determination which leads to the dominant uncertainty in the inferred age of
the oldest globular clusters.  Based on the distance estimators then
available, a 95\% lower limit of $12.1$ Gyr was determined for the
mean age of 17 of the oldest globular
clusters.
By comparison, the age of a flat-matter dominated universe with Hubble constant
$H_0 = 100 h$ km s$^{-1}$ Mpc$^{-1}$, is $6.51 [80/h]$ Gyr, The difference
between the lower and upper limits 
quantified the extent of the cosmic ``age problem".   This analysis was based
on normalizing the color-magnitude diagram by utilizing the inferred instrinsic
magnitude of RR-Lyrae stars at a mean metallicity of  [Fe/H] $= -1.9$ to fix
the magnitude of the horizontal branch for the globular clusters under study. 
By comparing the difference between this magnitude, and the magnitude of the
observed main-sequence turnoff point for these clusters to the predicted
magnitude difference based on stellar evolution modelling, one infers the age
of the clusters.  Other analyses obtained similar limits (i.e. see
\cite{bolte}).

Recently, following Hipparcos parallax measurements both for Cepheid variable
stars and for various subdwarf main sequence stars (i.e. \cite{feast,reid}), we
have been prompted to reanalyze the various results on the distance scale to
globular clusters, and the resulting age estimates \cite{chabetal3}.  The
changes have been dramatic.  A systematic shift in the estimated distances to
globular clusters has made it clear that the earlier apparent convergence of
such estimates was fortuitous, and the distance data was, and still remains,
dominated by systematic errors.  Our new best fit age is slightly lower than
our previous 95\% lower limit, and the lower limit we obtained (by accounting
for the now-apparent systematic uncertainties in the RR Lyrae distance
estimators) is $9.6$ Gyr.  Even allowing for minimum time of $0.2$ Gyr after
the BIg Bang for
the galaxy to form, this new lower limit is consistent with a flat matter
dominated universe Hubble age if $H_0 \le 66$.

While this might suggest that the need to consider a cosmological constant has
now vanished, it is important to remember that the age problem was just
one, albeit an important one, of several cosmological arguments which together
pointed in this direction. Thus one must combine this new age estimate with
the other constraints, which themselves have evolved, in order to 
reanalyze this issue.  Following
\cite{kraussturn}, I display in Figure 1a, for a flat Universe with
cosmological constant
$\Lambda$,   the parameter space of 
 $h$, vs $\Omega_0=
\rho_{matter}/\rho_{crit} $ (where $\rho$ is density and $\rho_{crit}$ is the
critical density for a flat universe, so that $\Omega_{\Lambda} = 1-
\Omega_0$), showing the allowed region given new age constraint, 
$ 15 \ge \tau \ge 9.8$ Gyr, along with the other new constraints I shall
describe momentarily.  For reasons which will also become clear shortly, I 
present in Figure 1b, the allowed range in the same parameter space for an open
universe.  

The quoted upper limit of 15 Gyr is obtained by taking the
$95 \% $ upper limit of $ \approx 14$ Gyr obtained from the analysis in
\cite{chabetal3} and adding 1 Gyr as an upper limit to the estimated time
after the big bang before star formation began in what would become the halo
of our galaxy. Based on recent observations of primeval galaxies at redshifts
in excess of 3, this seems a reasonable upper limit on the time before
structures began to form.

The other constraints displayed in figure 1 come from a consideration of the
two other independent fundamental sets of cosmological observables, the baryon
content of the universe, and large scale structure. I shall describe
each in turn below.   First, however, it is worth pointing out that current
estimates of the Hubble Constant have themselves evolved.  Two years ago,
there was apparent incompatibility between HST measurements based on Cepheid
distances to Virgo, and those based on Supernova Type 1a
distance measurements.   These two distance measures have been converging
(i.e. \cite{freedman}), so that now a range for 
$H_0$ of $ \approx 65 \pm 13$ brackets both measurements.  The horizontal dashed
lines in the figures display this presently preferred suggested range for
$H_0$. 

For over 20 years a robust upper limit on the baryon density 
of the universe has come from considerations of Big Bang Nucleosynthesis (BBN).
The predicted primordial abundance of the light elements up to $^4He$,
all of which are known to be produced primarily during BBN and not in stars, is
a function in each case of the baryon to photon ratio in the universe.  Hence a
detailed comparison of all the inferred primordial light element abundances
with predicted abundances yields a
restricted allowed range of baryon densities.  That is, of course if there is
concordance. 

During the past two years there has been a great deal of heat, and some
light, shed in this area.  First, starting about two years ago, BBN predictions
\cite{kk1} began to tighten the allowed range of $\Omega_B$.  It began to
become clear that unless systematic uncertainties in $^4He$ were allowed for
the upper limit on $\Omega_B$ coming from $^4He$ alone was becoming dangerously
close to the lower limit.  

The most robust upper limit on $\Omega_B$, however has for over 20
years come from observations of deuterium.  Deuterium is only destroyed in
stellar processing, so any observation of the interstellar or solar system
 abundance would put a lower bound on the actual primordial abundance
of deuterium. Since the predicted deuterium BBN remnant abundance is a
monotonically falling function of $\Omega_B$, a lower bound on deuterium
translates into an upper bound on $\Omega_B$.   The observed instellar value
of $D/H \ge (1.6 \pm 0.2) \times 10^{-5}$ puts a limit $\Omega_B h^2 \le .027$.

This situation took on a new dimension two years ago with the first claimed
observation of deuterium absorption lines in primordial hydrogen clouds
illuminated by distant quasars \cite{songaila}.  This method in principle
allows a direct determination of the primordial deuterium abundance, and
hence a direct measure of $\Omega_B$.  The only problem is that shortly
after the original measurement, which gave an anomalously high value for
the primordial deuterium abundance, other observations gave a value almost
one order of magnitude lower.  This value, $D/H = (2.4 \pm 0.3 \pm 0.3) \times
10^{-5}$ \cite{tytler1,tytler2}, is almost equal to the lower limit quoted
above, and suggests not only that $\Omega_B$ is near its upper limit, but also
that there has been little chemical evolution of $^2H$.  Moreover, it
reinforces the requirement that their be systematic errors in the $^4He$
abundance estimates if there is to be concordance in BBN.
\cite{KK3,copist,hata}.

Most recently, the original high deuterium observation has been
withdrawn.  However, at the present time, until more observations are made, it
seems premature to require that $\Omega_B$ is near its upper limit.  We have
seen over and over again that systematic errors are the dominant contribution
to our uncertainty in cosmological quantities, and thus large shifts within the
allowed range are as likely as small shifts. Moreover, concordence between the
deuterium estimates and the $^4He$ estimates requires some significant
systematic error in one of these values. Nevertheless, it is reassuring that
even with the new data, a robust estimate of allowed range of
$ 0.01 \le \Omega_B h^2 \le 0.0265$ \cite{KK3} from BBN, allowing for maximal
systematic uncertainties in all abundance estimates remains compatible with the
data.  This range is slightly larger than another
independent estimate
\cite{copist} but is more easily compatible with the new
deuterium abundance observations.

Even this relatively large upper bound on $\Omega_B h^2$ is clearly
incompatible with a flat baryon dominated for any reasonable Hubble Constant. 
What is more surprising, however, is that it seems incompatible with a flat
universe at all unless the Hubble constant is extremely small, or there
remains a large unclustered component of the energy density of the universe. 
This result arises from considerations of X-Ray measurements of hot gas in
rich galaxy clusters.  If the gas in these systems is in hydrostatic
equilibrium, and the overall gravitational potential is relativity smooth, one
can invert an X-Ray temperature/luminosity profile to get both the mass in hot
gas, which is by far the largest baryonic component to the mass of the
cluster, and also the total mass of the cluster.  Thus, one gets a direct
estimate for $ f_B=\Omega_B/\Omega_{0}$, where $\Omega_{0}$ represents the
fraction of the closure density which is in the form of clustered mass in the
universe. If one assumes, for example, that $\Omega =1$, and non-relativistic
matter dominates, then $\Omega_{0}=\Omega =1$.

In 1993, White {\it et al} \cite{whiteetal} argued that the estimate of $f_B$
obtained from the Coma cluster was sufficiently large so that it would be
incompatible with the BBN upper limit for a flat matter dominated universe
unless the Hubble constant were extremely small.  Since that time a number of
analyses of a broader class of clusters confirms the large $f_B$ estimates
\cite{whitefab,henry}, and these were utilized in previous work to put
constraints in
$\Omega_0$ and
$h$ space \cite{kraussturn,oststein}.   Recently, a 
comprehensive theoretical analysis of cluster modeling has been completed
\cite{evrard} to explore the robustness of the X-Ray interpretations.  There is
remarkable consistency between the numerical profiles and the data for a value
of $f_B = 0.060 \pm .003) h^{-3/2}$.   A comparison of this value with the BBN
constraint would put very severe limits on $\Omega_0$.  However, if one
desires a more conservative upper limit on $\Omega_0$ one should consider the
lowest possible value of $f_B$ which would be consistent with the data.  
This value assumes all baryons are in the form of hot gas in the
cluster and is given by $0.043 h^{-3/2}$ \cite{evrard}.  Similarly, taking
the largest value of $f_B$ consistent with all the  data, and allowing for a
baryon fraction as large as .013 from stars and dark baryonic halo objects
yields an approximate upper limit $0.078 h^{-3/2}$ for the range of $h$ values
of interest here.

Combining these limits with the BBN bounds yields a conservative 
constraint $
0.12
\le
\Omega_0 h^{1/2} \le 0.60 $.   This is displayed in Fig 1 for both an open and
flat universe with cosmological constant. 

Finally, the last set of constraints comes from considerations of large scale
structure.  This is another area where progress has been made
recently, in particular as a result of measurements of CMB anisotropies.
 For some time, various independent observations have suggested
that while the clustered mass in the universe exceeds the upper limit on
the baryon density coming from BBN, it nevertheless falls short of the
closure density.   Recent observations confirm this trend.  Nevertheless,
virial estimates suggest that 
$\Omega_0 \ge 0.3$, and I  adopt this conservative lower bound here, as
displayed in Fig 1.

I shall utilize two additional large scale structure constraints in
this analysis.  First, observations of galaxy correlations supply a constraint
on the shape of the power spectrum of density fluctuations on large scales. 
This is an extremely powerful constraint because it is related to the
primordial power spectrum primarily by considerations of causality, and the
density of clustered matter in the universe.  While the result is somewhat
model dependent, assuming a Cold Dark Matter dominated universe and assuming a
roughly scale invariant initial spectrum of density perturbations, as
predicted by inflation and as allowed by COBE constraints on the CMB, Peacock
and Dodds \cite{Pkdod} obtained the constraint $\Gamma =\Omega_0 h
exp(-\Omega_B -\Omega_B/\Omega_0) = 0.255^{+.038}_{-.033}$, which has been used
in previous analyses.  More recently both these authors \cite{pk, pkdod96} have
noted that non-linear effects might alter their results on short scales. 
Removing the shortest scale points and refitting one obtains
\cite{liddle1,liddlw}
$\Gamma = 0.230^{+.042}_{-.034} + 0.28(1/n -1)$ at the 95 per cent
confidence level.  Here n is the spectral index of the
primordial density perturbations, which CMB observations suggest is
between $ \approx 0.9-1.1$. 
We display this constraint in both Figure 1a and 1b, as it is insensitive to
the presence or absence of a cosmological constant. 

Finally, recently a number of authors \cite{vianaliddle1,lidvian2,others} have
examined the abundance of galaxy clusters, which constrains the magnitude of
density fluctuations on intermediate scales. By comparing to the COBE
normalized value on large scales \cite{bunnwhite}, one can put a constraint
on $\Omega_0$ vs $h$ which is complementary to the shape constraint described
above, although it turns out to be provide limits which are quite similar. 
Because the growth of fluctuations between COBE scales and galaxy cluster
scales is dependent on the geometry of the universe, this constraint is
slightly different for open vs flat cosmological constant dominated
cosmologies.  We plot these 
constraints \cite{liddle1,liddlw}
in Figures 1a and 1b.

The interplay between all of these new constraints in the
$\Omega_0$ vs
$h$ parameter space is quite significant.   Most important, by shifting the
allowed region in $h$ upward, the new age limit combines with the other
constraints so that now a {\it larger} region of parameter
space is allowed for an open universe than for a flat universe with
cosmological constant.  In particular, it is now clear that smaller values
of a Hubble constant yield a universe which is now {\it too old} in a
cosmological constant dominated universe.   The significance of the 
shift in age estimate which has taken place can be seen in both figures, where
a solid curve displays what used to the upper limit on the allowed phase
space coming from the earlier age constraint. For an open universe, this old
limit was extremely constraining. In this regard, the recent decrease in
estimates for the Hubble constant coming from HST measurements is also
relevant. 

It is also clear that while globular cluster age estimates have relaxed the
constraint on the overall matter density of the universe, the other
updated cosmological constraints coming from large scale structure
and baryon counting have solidified in a region in which a flat,
matter dominated universe is marginally viable only for an extremely small
Hubble Constant.

The debate thus appears to remain between an open universe and flat universe
with cosmological constant.  On the basis of shear size of allowed
parameter space, for the first time an open universe is more strongly favored.
There are, however,  additional theoretical and observational
data which bear on this conclusion.  First, several preliminary 
observations have been made which are claimed to 
directly constrain $\Omega_0$
and
$\Lambda$ which on balance may swing slightly in favor of a non-zero
cosmological constant.   Cluster evolution, in number density \cite{bahc}, and
luminosity density\cite{sato}, has recently been used to place
constraints on both
$\Omega_0$, and $\Lambda$.  The former, also displayed in figures 1a and 1b,
yields a bound on $\Omega_0$ which is inconsistent with a flat universe,
but reasonably independent of the presence or absence of a cosmological
constant.   The latter however, is claimed to
strongly favor a non-zero cosmological constant.  In fact, 
a {\it lower bound} on $\Omega_{\Lambda} $ of 0.37 at the $99 \%$ confidence
level is claimed.  This result is displayed in figure 1a, but it should be
taken as preliminary.  To counterbalance it, recent observations
of the luminosity-redshift relation for type 1a supernovae \cite{perlm} place a
preliminary {\it upper limit} on the value of $\Omega_{\Lambda} $ of 0.49. 
While this result involves
the first application of a new technique, the recent measurements of more high
redshift supernovae should, within the next few years, provide the
strongest constraints on a cosmological constant in the universe today. 

Eventual measurements of cosmic microwave
anisotropies on small angular scales will allow a direct measurement of the
cosmological constant, in addition to other fundamental cosmological
parameters, at the 10-20 $\%$ level (i.e. \cite{kosowsky}).  A distinction can
be made between the effects of a cosmological constant, and a non zero
curvature.  For a flat universe, increasing $\lambda$, which implies decreasing
the ratio
$\Omega_0/\Omega_B$, causes the magnitude of the first Doppler peak to
increase, while its position is relatively unaffected.  An open universe, on
the other hand, changes the position of the first doppler peak, because
introducing a non-zero curvature alters the redshift-angular scale relation of
the universe.   In this regard, present measurements of CMB anisotropies
suggest a large doppler peak in the location expected for a flat universe. 
Thus, while extremely preliminary, CMB measurements on small angular scales
may favor of a cosmological constant over an open
universe \cite{krscwh}.

At present, it is clearly too early to choose one
cosmological model over the other. It is clearly getting increasingly
difficult to find accord with a flat universe without a cosmological
constant.  The question then becomes: Which fundamental fine tuning problem is
one more willing to worry about: the flatness problem, or the cosmological
constant problem?  The latter involves a fine tuning of almost 125
orders of magnitude, if the cosmological constant is non-zero and comparable
to the density of clustered matter today, while the former involves a fine
tuning of perhaps only 60 orders of magnitude if one arbitarily fixes the
energy density of the universe at the planck time to be slightly less than
the closure density.  Numerological arguments might thus suggest that one
should be more prepared to give up flatness than a zero cosmological constant.
I claim however, that this argument is incomplete.

We have a perfectly good theory, involving physics well below the planck
scale, for why we might live in a flat universe.   As long as there was an
inflationary regime in the early universe, the universe generically is driven
to be approximately flat to many more decimal places than are required to
resolve the flatness problem.  Moreover, when considering possible particle
physics models of the early universe, inflation seems to be ubiquitous.  The
difficulty seems to be not how to get enough inflation, but rather how to end
it.   On the other hand, we have absolutely no theory of the cosmological
constant at all.  Other than vague {\it a priori} prejudice, there is no well
defined physical argument at the present time suggesting a zero,
rather than arbitrarily small value of this quantity.  Moreover, the energy
scale associated with a non-zero cosmological constant which dominates the
universe today is not unusual. It corresponds to the characteristic mass scale
which is discussed for neutrino masses which might solve the solar neutrino
problem.  Also interesting, recent arguments suggest that if the laws of
physics predict a distribution of universes, with randomly chosen
values for the cosmological constant, then quantitative anthropic arguments
make it not implausible that it should be observed to be comparable to the
matter density in the universe today \cite{weinberg}.  Whatever one's views
toward anthropic arguments, it is not clear that the same
reasoning could be applied to the flatness problem.  Precisely because we have
physical laws which suggest the universe should be flat, I would argue the
{\it a priori} probability distribution for the curvature parameter one
might reasonably consider should be strongly peaked about zero, in which case
anthropic arguments along the lines applied to the cosmological constant might
not be as suggestive, to the extent such arguments are suggestive.

In conclusion, we should know within a decade whether the cosmological
constant is non-zero, and whether we live in a flat universe.  The recent
resolution of the age problem has dramatically altered the case for
a cosmological constant vs an open universe. A combination of cosmological
observations now allows a larger parameter space for an open universe than a
flat universe with cosmological constant. While one might argue that
theoretical prejudice still favors the latter, if either of these
cases represents reality, the implications for
fundamental particle physics will be profound.

\section*{Acknowledgments}
This work was supported in part by the DOE, and Case Western Reserve
University.  I wish to thank CERN for a stimulating environment while this work
was completed, my collaborators on globular cluster age estimation, Brian
Chaboyer, Pierre Demarque, and Peter Kernan, and Steve Weinberg for
enlightening discussions on anthropic arguments in favor of a cosmological
constant.

\newpage
\begin{figure}[htb]
\vglue 4.0in
\includegraphics{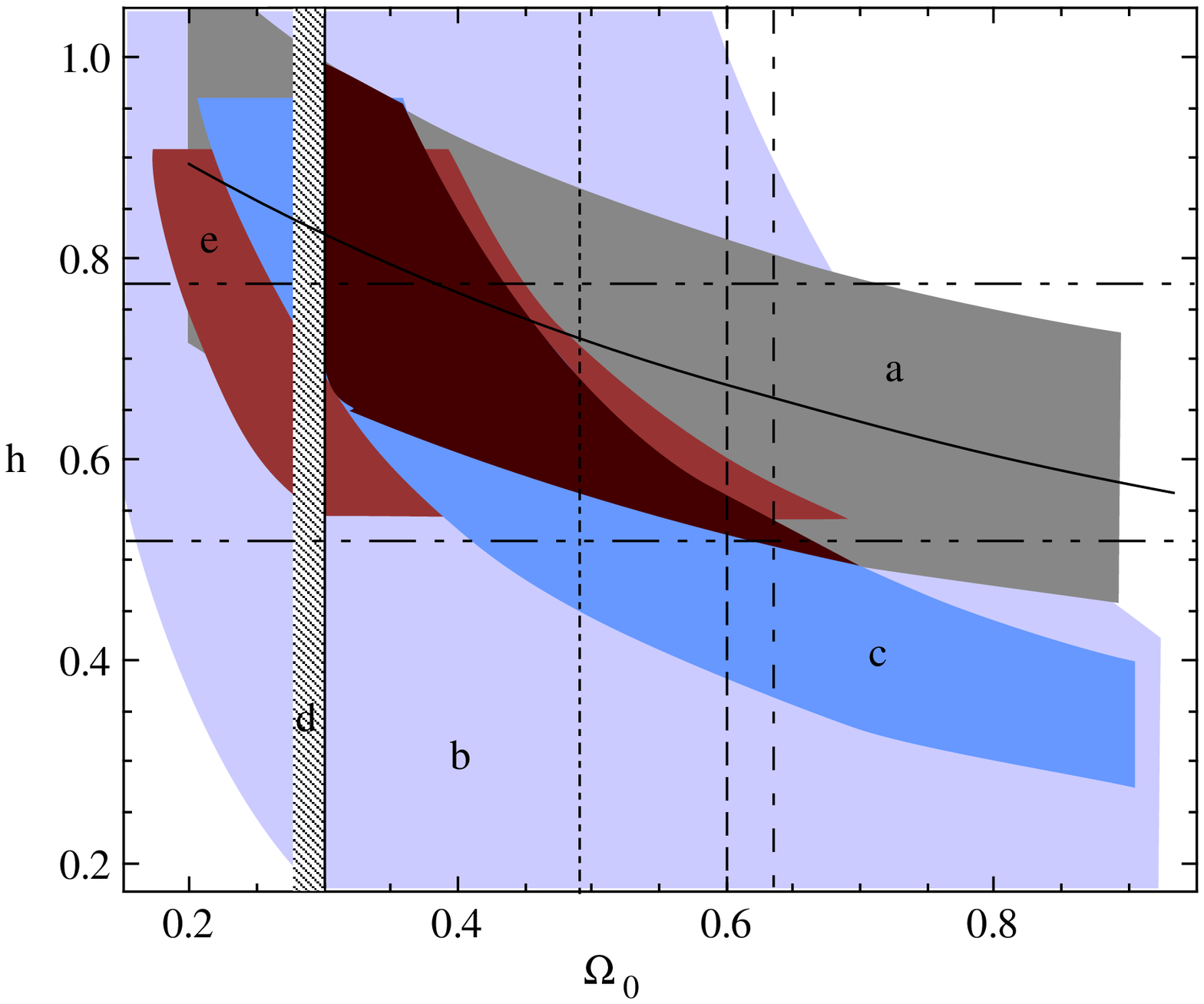}
\caption{ Shaded regions represent constraints on $h$ vs. $\Omega_0$ for a flat
universe with cosmological constant arising from (a) globular cluster age
limit, (b) baryon content of the universe, (c)shape of galaxy power spectrum,
(d) lower limit on clustered mass from virial estimates, (e) abundance of
galaxy clusters extrapolating from COBE normalization, assuming dark matter is
cold. Shown in dark shading is the locus of points in phase space satisfying
all limits.   Horizontal dashed lines present the upper and lower limits of
the present preferred range of the Hubble constant quoted in the text.  The
dashed vertical lines represent various preliminary limits on $\Omega_0$ for a
flat universe with cosmological constant.  The right-most limit represents the
claimed $99\%$ lower limit on $\Omega_\Lambda=0.34$ arising from
considerations of the evolution of the cluster luminosity function.  The next
largest limit represents an upper limit on $\Omega_0$ from considerations of 
the evolution of the galaxy cluster number density. The left most
limit arises from a claimed upper limit on $\Omega_\Lambda$ coming from a
measurement of $q_0$ using Type 1a supernovae.  Finally, the solid curved line
represents the previous upper limit on $h$ vs $\Omega_0$ using the
claimed lower limit on the age of the Universe of 12.1 Gyr which has now been
revised downward. }
\label{figcosmo}
\end{figure}

\begin{figure}[htb]
\vglue 2.8in
\includegraphics{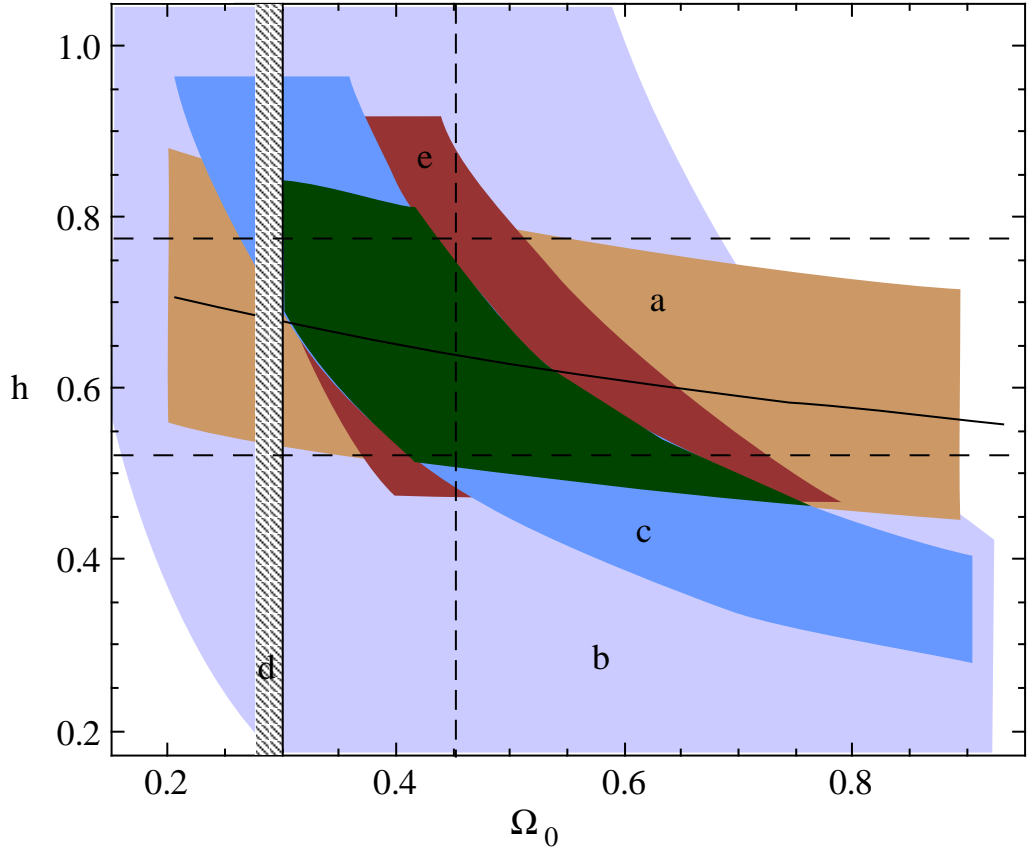}
\caption{  Same as figure 1a, but in this case for an Open Universe. The
vertical dashed line represents  the claimed upper limit on
$\Omega_0$ arising from considerations of the evolution of the galaxy cluster
number density.}
\label{figopen}
\end{figure}

\end{document}